\documentclass[aps,prd,preprintnumbers,superscriptaddress,showpacs,nofootinbib]{revtex4}
\usepackage{amssymb,amsmath}
\usepackage{bm}
\usepackage{slashed}
\usepackage{graphicx}
\usepackage{epstopdf}
\usepackage{tabularx}
\usepackage{float}

\usepackage{textcomp}

\usepackage{color}
\usepackage[colorlinks, hyperindex]{hyperref}
\hypersetup{
  colorlinks,
  citecolor=blue,
linkcolor=red,
urlcolor=blue,
}

\begin{document}

\title{Relativistic chiral description of the $^1S_0$ nucleon-nucleon  scattering }

\author{Xiu-Lei~Ren}
%\email[]{Email: xiulei.ren@pku.edu.cn}
\affiliation{State Key Laboratory of Nuclear Physics and Technology, School of Physics, Peking University, Beijing 100871, China}
\affiliation{Institut f\"{u}r Theoretische Physik II, Ruhr-Universit\"{a}t Bochum, D-44780 Bochum, Germany}

\author{Chun-Xuan Wang}
%\email[]{Email: kaiwen.li@buaa.edu.cn}
\affiliation{School of Physics, Beihang University, Beijing 102206, China}

\author{Kai-Wen~Li}
%\email[]{Email: kaiwen.li@buaa.edu.cn}
\affiliation{School of Physics, Beihang University, Beijing 102206, China}
\affiliation{Yukawa Institute for Theoretical Physics, Kyoto University, Kyoto 606-8502, Japan}

\author{Li-Sheng~Geng}
\email[]{Email: lisheng.geng@buaa.edu.cn}
\affiliation{School of Physics, Beihang University, Beijing 102206, China}
\affiliation{Beijing Key Laboratory of Advanced Nuclear Materials and Physics, Beihang University, Beijing 100191, China}
\affiliation{School of Physics and Microelectronics, Zhengzhou University, Zhengzhou, Henan 450001, China }

\author{Jie~Meng}
\email[]{Email: mengj@pku.edu.cn}
\affiliation{State Key Laboratory of Nuclear Physics and Technology, School of Physics, Peking University, Beijing 100871, China}
\affiliation{School of Physics, Beihang University, Beijing 102206, China}
\affiliation{Yukawa Institute for Theoretical Physics, Kyoto University, Kyoto 606-8502, Japan}
%\affiliation{Department of Physics, University of Stellenbosch, Stellenbosch 7602, South Africa}

\date{\today}

\begin{abstract}
Recently, a relativistic chiral nucleon-nucleon interaction was formulated up to leading order which provides
a  good description of the phase shifts of $J\leq1$ partial waves [Chin.Phys. C42 (2018) 014103]. Nevertheless, a separable regulator function that is not manifestly covariant was used in
solving the relativistic scattering equation. In the present work, we first explore a covariant and separable form factor to regularize the kernel potential and  then
apply it to study the simplest but most challenging $^1S_0$ channel which features several low-energy
scales. In addition to being self-consistent,
we show that the resulting relativistic potential can describe quite well
the unique features of the $^1S_0$ channel  at leading order,
in particular  the pole position of the virtual bound state
and the zero amplitude at the scattering momentum $\sim 340$ MeV, indicating that
the relativistic formulation might be more natural from the point of view of effective field theories.

\end{abstract}

\maketitle

\section{Introdcution}

Chiral nuclear forces have been  studied
extensively~\cite{Bedaque:2002mn,Epelbaum:2008ga,Machleidt:2011zz,Epelbaum:2012vx}  since Weinberg extended chiral
perturbation theory (ChPT)~\cite{Weinberg:1978kz}, an effective field theory (EFT) of low-energy QCD, to describe
nucleon-nucleon~({\it NN}) scattering in the 1990s~\cite{Weinberg:1990rz,Weinberg:1991um}. Using the Weinberg power counting, two famous chiral forces were constructed up to next-to-next-to-next-to-leading order a decade ago~\cite{Entem:2003ft,Epelbaum:2004fk}, which have been applied in the {\it ab-initio} descriptions of nuclear structure and reactions~(see e.g. Refs.~\cite{Epelbaum:2011md,Tews:2012fj,Epelbaum:2012iu,
Hergert:2012nb,Hergert:2013vag,Epelbaum:2013paa,Jansen:2014qxa,Bogner:2014baa,Lynn:2014zia,Hagen:2015yea,Elhatisari:2015iga, Lapoux:2016exf,Elhatisari:2017eno}).
Recently, the Bochum-J\"{u}lich and Indaho groups have constructed chiral forces at the fifth order and  shown that the
resulting
description of {\it NN} phase shifts and scattering
data~\cite{Epelbaum:2014sza,Reinert:2017usi,Entem:2014msa,Entem:2017gor} is comparable to the high precision
phenomenological nuclear potentials~(such as Reid93~\cite{Stoks:1994wp}, Argonne $V_{18}$~\cite{Wiringa:1994wb}, and
CD-Bonn~\cite{Machleidt:2000ge})
with a $\chi^2/\mathrm{datum}\sim 1$.

On the other hand, with the developments of covariant density functional theories
and covariant chiral
perturbation theory,
 relativistic effects are found to play an important role in nuclear structure~\cite{Mengbook} and
one-baryon and heavy-light systems~\cite{Geng:2008mf,Ren:2012aj,Ren:2014vea,Altenbuchinger:2013vwa} (for a short
review see Ref.~\cite{Geng:2013xn}).
Therefore, we proposed  to construct a relativistic
nuclear force and baryon-baryon interactions in covariant ChPT~\cite{Ren:2016jna,Li:2016mln,Song:2018qqm,Li:2018tbt,Xiao:2018jot,Bai:2020yml,Xiao:2020ozd,Song:2020isu,Liu:2020uxi}, which would provide essential inputs for relativistic many-body calculations,
such as the Dirac-Brueckner-Hartree-Fock theory~\cite{Brockmann:1990cn,Shen:2016bva,Shen:2017vqr,Shen:2019dls}.
In Ref.~\cite{Ren:2016jna}, we explored a covariant power counting of {\it NN} scattering  at
LO and found that a good description of phase shifts of angular momentum $J=0,~1$  can be achieved by solving the
relativistic three-dimensional reduction of the Bethe-Salpeter equation~\cite{Salpeter:1951sz}, i.e. the Kadyshevsky
equation~\cite{Kadyshevsky:1967rs}.
Since the intermediate momentum in the scattering equation runs from zero to infinity,
the kernel potential must be damped by a regulator function at high momentum to avoid divergence.
Such regulator functions are usually referred to as form factors (FFs).
In principle, the choice of FFs is rather  {\it ad hoc}~\cite{erkelenz:1974uj} and the impact on physical observables should be removed or minimized.

In the literature, there exist several kinds of
form factors, such as the following one
\begin{equation}
   f(\bm{q}^2) = \frac{\Lambda^2}{\Lambda^2 + \bm{q}^2},
 \end{equation}
given by Ueda~\cite{Ueda:1969er}, where $\bm{q}$ is the three-momentum transfer, and $\Lambda$ denotes the
cutoff. In Ref.~\cite{Machleidt:1987hj}, the Bonn potential was constructed with monopole ($n=1$) or dipole ($n=2$) FFs
\begin{equation}
   f(\bm{q}^2) = \left[\frac{\Lambda^2-m_\phi^2}{\Lambda^2-\bm{q}^2}\right]^n,
 \end{equation}
 for different meson ($\phi$)-nucleon-nucleon vertices. In Refs.~\cite{Jackson:1975be,Holinde:1976qwa}, an
   eikonal form facor~\cite{Woloshyn:1971zz} was used to construct the one-boson-exchange potentials.
 A Gaussian FF,
\begin{equation}
   f(\bm{q}^2) = \mathrm{exp} \left[ -\frac{\bm{q}^2}{\Lambda^2} \right],
 \end{equation}
 was firstly employed by the Nijmegen group~\cite{Nagels:1977ze} and then applied in the study of chiral forces~\cite{Ordonez:1993tn}.
 In Ref.~\cite{Epelbaum:1999dj} , Epelbaum {\it et al.} proposed a seperable form factor (SFF),
 \begin{equation}\label{Eq:SFF}
   f(\bm{p}) = \mathrm{exp}\left[-\left( \frac{\bm{p}^2}{\Lambda^2} \right)^n \right],
 \end{equation}
 which only depends on the initial (final) three-momenta $\bm{p}$ ($\bm{p}'$). In comparison with the
 $\bm{q}^2$-dependent FFs, which
 introduce additional angular dependence to partial
  wave potentials and thus affect the interpretation of contact interactions of chiral nuclear forces~\cite{Nogga:2005hy,Epelbaum:2000kv}, this separable form factor is better suited in constructing chiral forces~\cite{Entem:2003ft,Epelbaum:2004fk,Entem:2014msa}.
Recently, a regulator function more proper  for the long-range interaction in coordinate space was proposed~\cite{Epelbaum:2014efa}.

In our previous study~\cite{Ren:2016jna}, we took the SFF to regularize the kernel of the relativistic LO chiral
potential.
However, this form factor is not covariant. In order to maintain the self-consistency, a covariant FF is favored in the relativistic framework.
Therefore, in this work, we explore a separable form factor which is manifestly covariant.

As an application, we study the simplest but most challenging $^1S_0$ channel with this new form
factor. There are several particular features in the $^1S_0$ wave, such as the large variance of phase
shifts from $60^\circ$ to $-10^\circ$ with the laboratory energy $T_\mathrm{lab.}\leq 300$ MeV,
a significantly larger scattering length ($a=23.7$ fm) than the
pion Compton wave length, the zero amplitude, namely the zero phase shift of
$^1S_0$ with the center of mass (c.m.) momentum $k_0\sim 340$ MeV, a virtual bound state
around $i\gamma= -i10$~MeV. In Ref.~\cite{vanKolck:1998bw}, van Kolck  pointed out that a different kind of fine-tuning is needed to produce the zero
amplitude of $^1S_0$ in contrast with the  virtual bound state.
Since these typical energy scales, such as $-i10$ MeV and $340$ MeV~\footnote{In Ref.~\cite{Lutz:1999yr}, the chiral
  expansion of {\it NN}
potential is performed around the scattering momentum of the zero amplitude, $k_0=340$ MeV.}, are smaller than the chiral symmetry breaking
scale ($\Lambda_\chi\sim 1$ GeV), they should
be roughly reproduced at the lowest order according to the principle of effective field theories, as explored in
Ref.~\cite{SanchezSanchez:2017tws}.
Thus, the description of these quantities may be considered as a criterion to test a
natural power counting for the {\it NN} interaction. Inspired by Ref.~\cite{SanchezSanchez:2017tws}, where the unique features of $^1S_0$ are well described
simultaneously by rearranging the short-range interactions in non-relativistic ChPT,
we extend our previous work~\cite{Ren:2016jna} to perform a systematic study of the $^1S_0$ channel up to leading
order in the
relativistic framework
to describe/predict the related quantities with the covariant and separable form factor.

The manuscript is organized as follows. We first present the  $^1S_0$ potential in the relativistic formulation, and compare the covariant FF with
the non-covariant one. In Sec.III, we show the description of the $^1 S_0$ phase shifts and the predicted low energy quantities, followed by a short summary in Sec.IV.

\section{Theoretical Framework}

The relativistic chiral force is formulated up to LO in Ref.~\cite{Ren:2016jna},
where the  $^1S_0$ potential reads
\begin{eqnarray}\label{Eq:1S0V}
  V_{1S0}(p', p)&=&4\pi C_{1S0} + 2\pi (C_{1S0}+\hat{C}_{1S0}) \frac{E_pE'_p-m_N^2}{m_N^2}\nonumber\\
  && + \frac{\pi g_A^2}{f_\pi^2}\int^1_{-1}dz \frac{1}{q^2-m_\pi^2+i\epsilon} \left(E_pE_{p'}- p \, p'\, z-m_N^2\right),
\end{eqnarray}
with the axial vector coupling $g_A=1.267$ and the pion decay constant $f_\pi=92.4$ MeV~\cite{Olive:2016xmw}.
The four vector $q$ represents the momentum shift with $q^0=(E_{p'}-E_{p})$ and $\bm{q}=\bm{p}'-\bm{p}$,
$\bm{p}$ ($\bm{p}'$) is the spatial component of initial (final) momentum of the nucleon in the center of mass (c.m.) frame
with $E_p=\sqrt{\bm{p}^2+m_N^2}$ ($E_{p'}=\sqrt{\bm{p}'^2+m_N^2}$), $p$ and $p'$ are defined as the magnitudes of the momenta, $p=|\bm{p}|$, $p'=|\bm{p}'|$, and $z$ denotes the cosine of the angle between $\bm{p}$ and
$\bm{p}'$.
It
should be noted that there are two unknown parameters, $C_{1S0}$ and $\hat{C}_{1S0}$, which are the combinations of the five
LECs, $C_S$, $C_A$, $C_V$, $C_{AV}$, $C_T$, appearing in the lowest order chiral $NN$
Lagrangian~\cite{Polinder:2006zh,Djukanovic:2007zz} with $C_{1S0} =C_S+C_V+3C_{AV}-6C_T$ and $\hat{C}_{1S0}=3C_V
+C_A+C_{AV}-6C_T$.

Because the nuclear force is non-perturbative, one has to resum the above potential via a scattering equation, such as the
Bethe-Salpeter equation~\cite{Salpeter:1951sz} or its three-dimensional reductions~\cite{Woloshyn:1974wm}. In Ref.~\cite{Ren:2016jna}, we employed the Kadyshevsky
equation~\cite{Kadyshevsky:1967rs}~\footnote{Our numerical results would remain almost the same, if the Thompson equation~\cite{Thompson:1970wt} and the Blankenbecler-Sugar
  equation~\cite{Blankenbecler:1965gx} were employed instead.} in the
c.m. frame, which reads
\begin{equation}\label{eq:kadyshevsky}
   T_{1S0}(p', p|W) = V_{1S0}(p', p|W) + \int_0^{+\infty}\frac{k^2d k}{(2\pi)^3} V_{1S0}(p', k|W)~\frac{m_N^2}{2E_k^2(E_p-E_k + i\epsilon)} T_{1S0}(k, p|W),
 \end{equation}
  for the $^1S_0$ partial wave,
 where $W=(\sqrt{s}/2, \bm{0})$ is half of the total four-momentum with the total energy $\sqrt{s}=2E_p=2E_{p'}$.
In solving the scattering equation, one has to use a form factor to regularize the kernel
potential as mentioned in the introduction,
\begin{equation}
  V_{1S0}(p, p'|W) \rightarrow f(\bm{p}|W)~ V_{1S0}(p, p'|W)~ f(\bm{p}'|W).
\end{equation}

\begin{figure}[b]
\centering
  \includegraphics[width=10cm]{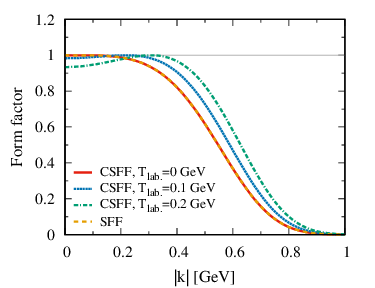}\\
\caption{Form factors as a function of momenta with the cutoff $\Lambda=600$ MeV. The solid, dotted, and dot-dashed  lines represent the CSFF for
  $T_\mathrm{lab.}=0,~100,~200$ MeV, respectively. The dashed line denotes the SFF.}
\label{Fig:CSFF}
\end{figure}

Here, we  introduce
  a covariant and separable form factor (CSFF), which has the exponential form
  \begin{equation}
    f^\mathrm{CSFF}(\bm{p}|W)
    =  \mathrm{exp}\left[-\left(\frac{\frac{s}{4} - \bm{p}^2 -m_N^2}{\Lambda^2}\right)^2\right].
  \end{equation}
  One can see that CSFF is
 trivial (equal to unity) for on-shell potentials with $\bm{p}^2= E_p^2 - m_N^2$. While, for the half-/full-off shell potentials appearing
 in the scattering equation, this form factor becomes
 \begin{equation}
 f^\mathrm{CSFF}(\bm{k}|W)
    =  \mathrm{exp}\left[-\left(\frac{\frac{1}{2} m_N T_\mathrm{lab.} - \bm{k}^2 }{\Lambda^2}\right)^2\right],
  \end{equation}
  where $T_\mathrm{lab.}$ is the laboratory kinetic energy with $s=4m_N^2 + 2m_N T_\mathrm{lab.}$.
The behavior of the CSFF as a function of $|\bm{k}|$ is shown in Fig.~\ref{Fig:CSFF} with $T_\mathrm{lab.}=0,~100,~200$
MeV, and the cutoff $\Lambda$ is fixed at $600$ MeV.
For
comparison, the SFF [Eq.~(\ref{Eq:SFF})] with $n=2$ is also given.
One can see that for $T_\mathrm{lab.}=0$, the SFF and CSFF are the same. While for $T_\mathrm{lab.}>0$,
  the CSFF is not monotonically decreasing in constrast to the SFF.
  The CSFF is smaller than the SFF and slightly increases to $1.0$ for $|\bm{k}|$
 between $0$ and $\sqrt{1/2 m_N T_\mathrm{lab.}}$. For $|\bm{k}|>\sqrt{1/2 m_N T_\mathrm{lab.}}$, the CSFF is larger
 than the SFF and  decreases to zero for $|\bm{k}|$ approaching $1$ GeV.

\section{Results and Discussion}

\begin{figure}[b]
\centering
  \includegraphics[width=10cm]{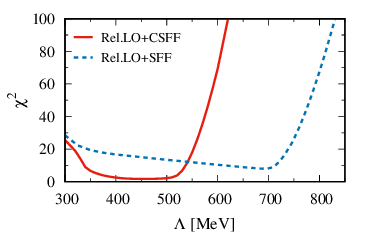}\\
\caption{Description of the $^1 S_0$ phase shifts, $\chi^2$, as a function of the cutoff $\Lambda$. The red solid and blue dashed lines denote the
  relativistic LO results
   obtained with the CSFF and the SFF, respectively.}
\label{Fig:La-chi2}
\end{figure}

In order to fine-tune the $^1S_0$ potential, the two parameters, $C_{1S0}$ and $\hat{C}_{1S0}$, are determined by fitting to the six data points of  the Nijmegen phase shifts for laboratory energy
$T_\mathrm{lab.}\leq 100$ MeV, as done in Ref.~\cite{Ren:2016jna}.
To calculate the phase shifts, the covariant and separable
form factor is employed to regularize the LO chiral potential and to
solve the Kadyshevsky equation.
We find that the minimum $\chi^2$, $\sum\limits_{i=1}^{6} (
\delta_i^\mathrm{LO} - \delta_i^\mathrm{Nij})^2$, is $1.64$ for $4$ degrees of freedom
when the cutoff is $460$ MeV. For comparison, we also use the SFF, as in Ref.~\cite{Ren:2016jna}, to describe the $^1S_0$ phase
shifts. The best fit result, $\chi^2=7.86$, locates at $\Lambda=695$ MeV. This shows that
different types of form factors could affect the phase shifts considerably.
Particularly, in our relativistic
framework, a self-consistent CSFF achieves a rather good description of the $^1S_0$ phase shifts.
In Fig.~\ref{Fig:La-chi2},
the evolution of fit-$\chi^2$ is shown with the cutoff changing from $300$ MeV to $850$ MeV.
One can see that the description of phase shifts is almost the same for the two FFs when the cutoff is around $300$ MeV. As $\Lambda$ increases, the fit-$\chi^2$ with the CSFF decreases more quickly than its counterpart with the SFF. The CSFF result shows a plateau, $\chi^2\sim 2.0$  with $\Lambda$ ranging from 350 MeV to 500 MeV.~\footnote{In principle, a small cutoff is favored to avoid the appearance of deeply bound states when one applied chiral nuclear forces to perform many-body
   studies.}
On the other hand, as the cutoff increases beyond $550$ MeV, the CSFF-$\chi^2$ becomes larger and increases faster than the SFF-$\chi^2$, which smoothly approaches to its minimum at $\Lambda = 695$ MeV and then starts to increase.~\footnote{The sharp increase of the $\chi^2$ with increasing cutoff can be traced back to the Wigner bound~\cite{Wang:2020myr}. In plain words, it means that with only contact  interactions, one cannot simultaneously reproduce the scattering length and effective range (or the phase shifts) of the $^1S_0$ channel with a large cutoff.}

\begin{figure}[t]
  \centering
  \includegraphics[width=8.5cm]{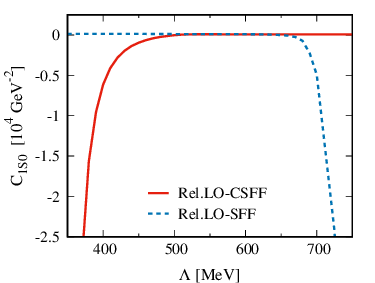}~~~~~
  \includegraphics[width=8.5cm]{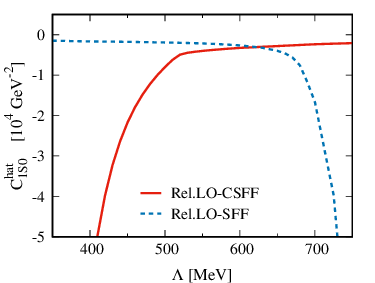}\\
  \caption{Evolution of the LECs, $C_{1S0}$ and $\hat{C}_{1S0}$, as a function of the cutoff $\Lambda$ with the CSFF and the traditional SFF.}.
  \label{Fig:1S0LECs}
\end{figure}

In Fig.~\ref{Fig:1S0LECs}, the corresponding LECs  $C_{1S0}$ and $\hat{C}_{1S0}$, obtained with the CSFF and the SFF, are plotted as
a function of the cutoff $\Lambda$. For the CSFF case, the magnitudes of both couplings decrease smoothly with the cutoff $\Lambda$ and do not show any special structure. On the other hand, the results obtained with the SFF show opposite trends. It is interesting to note that the coupling constants obtained  with both kinds of FFs agree more or less with each other in the cutoff range of 500$\sim$600 MeV.
Furthermore, the large difference between the couplings with the CSFF and those with the SFF for small cutoffs are related to the fact that the largest difference between both form factors appear for small cutoffs. On the other hand, the sharp decrease of the couplings with the SFF around $\Lambda=700$ MeV indicates the emerging impact of the Wigner bound, as can be seen in Fig.~2, i.e., the sharp increase of the fit $\chi^2$. On the other hand, for the CSFF case, the stabilization of the couplings around $\Lambda=550$ MeV indicates the emerging impact of the Wigner bound, as corroborated also by Fig.~2.

\begin{figure}[t]
  \centering
  \includegraphics[width=10cm]{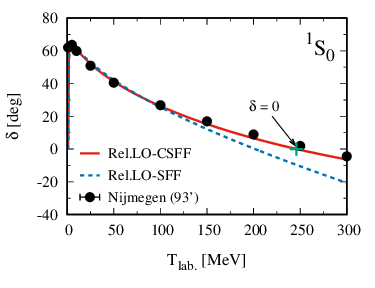}\\
  \caption{$^1S_0$ phase shifts $\delta$ as a function of laboratory energy $T_\mathrm{lab.}$. The red solid line
    denotes the best fitting results from the relativistic chiral NN potential with the covariant and separable form
    factor, while the blue dashed line is the result with the non-covariant form factor. The cross sign denotes the zero phase shift.  Solid dots represent
the  Nijmegen $np$ phase shifts~\cite{Stoks:1993tb}.
  }\label{Fig:1S0des}
\end{figure}

Next, we take the LECs from the best fit to plot the phase shifts up to
$T_\mathrm{lab.}=300$ MeV  in Fig.~\ref{Fig:1S0des}.
We see that both strategies yield a similar description of phase shifts with $T_\mathrm{lab.}\leq 100$ MeV.
Especially, at the very low energy region ($T_\mathrm{lab.}<1$ MeV), the phase shifts show a drastic increase from
$0^\circ$ to $62^\circ$, which is the same as the data of the NN-online database~\cite{nnonline}.
For the $T_\mathrm{lab.}>100$ MeV region, the calculated phase shifts with the CSFF agree better with the Nijmegen results
in comparison with the SFF phase shifts. It is interesting to note that the zero amplitude is
well reproduced around the c.m. momentum $k\sim 340$ MeV.

\begin{table}[t]
  \centering
  \caption{Predicted effective range expansion coefficients of the $^1S_0$ phase shifts (from the best fit results mentioned in the text) in comparison with the values from the Nijmegen PWA~\cite{Stoks:1994wp,PavonValderrama:2005ku}.}
  \label{Tab:Errpredc}
%\begin{ruledtabular}
\begin{tabularx}{16cm}{@{\extracolsep{\fill}}l|c|ccccc|c}
  \hline\hline
  &  $\Lambda$ [MeV]  &  $a$ [fm]  & $r$ [fm]     &  $v_2$ [fm$^3$]  &    $v_3$ [fm$^5$] &   $v_4$ [fm$^7$] & $i\gamma$ [MeV] \\
      \hline
  Nijmegen PWA
  &  --  &  $-23.7$   & $2.70$  & $-0.50$   & $4.0$ & $-20$   & $-i 10.0$  \\
  \hline
  Rel.LO-CSFF & $460$  & $-23.0$ & $2.61$ & $-0.66$ & $5.5$ & $-32$ & $-i 8.2$\\
  Rel.LO-SFF & $695$   & $-22.0$ & $2.53$ & $-0.75$  & $5.9$ & $-34$ & $-i8.5$ \\
  \hline\hline
\end{tabularx}
%\end{ruledtabular}
\end{table}

In addition, we predict the coefficients of the effective range expansion of the $^1S_0$ phase shifts,
\begin{equation}
  p\cot[\delta_{1S0}(p)] = -\frac{1}{a} + \frac{1}{2}r p^2 + v_2p^4 + v_3p^6 + v_4p^8 + \cdots,
\end{equation}
where $a$, $r$, and $v_{2,3,4,...}$ denote the scattering length, effective range, and curvature parameters,
respectively. Their values are tabulated in Table~\ref{Tab:Errpredc} in comparison with the results of the Nijmegen
partial wave analysis (PWA) extracted from the NijmII potential~\cite{Stoks:1994wp,PavonValderrama:2005ku}.
One can see that both the scattering length and effective range agree well with the Nijmegen PWA with a deviation of
about 5\%. On the other hand, the discrepancy of the shape parameters $v_{2,3,4}$ is a little bit large, which may be due to the lack of  two-pion exchange contributions in the present study~\cite{Epelbaum:2012ua}.

We also predict the binding momentum, $\gamma$, of  the virtual bound state in the $^1S_0$ channel,
which can be determined by searching
for poles in the scattering amplitude,
\begin{equation}
  p_B \cot[\delta_{1S0}(p_B)] \equiv i\, p_B,
 \end{equation}
 with $p_B=|\bm{p}_B|=i\gamma$ below threshold. In Table~\ref{Tab:Errpredc},
 the pole position of the virtual bound state is given. One can see that our result agrees with the empirical value.

We note in passing that a simultaneous fit of the partial wave phase shifts  with angular momentum $J\leq 1$ using the covariant form factor
yields a $\chi^2/\mathrm{d.o.f.} \approx 10$, slightly larger than that of Ref.~\cite{Ren:2016jna} obtained with the traditional separable form factor.~\footnote{However, the small difference between the fits with the CSFF and SFF should be better viewed as systematic uncertainties for the present study} Nevertheless,
the description of the $^1 S_0$ partial wave phenomena remains similar.

Finally, it should be noted that our current results are still cutoff dependent, as shown in
Fig.~\ref{Fig:La-chi2}. This unwelcome phenomenon was first noticed around 2000s~\cite{Kaplan:1998tg, Nogga:2005hy},
and even today remains an open question.
There are many works trying to construct a cutoff independent chiral {\it NN} potential by modifying the Weinberg power
counting, see e.g.,  Refs.~\cite{Birse:2005um,Timoteo:2005ia,Birse:2007sx,Yang:2007hb,Yang:2009kx,Yang:2009pn,
  Valderrama:2009ei,Valderrama:2011mv,Long:2011qx,Long:2012ve,Epelbaum:2012ua,Epelbaum:2015sha,SanchezSanchez:2017tws,Baru:2019ndr,Ren:2019qow}.
It will be interesting to investigate whether one can achieve cutoff independence in our relativistic framework in the future.

\section{Summary}
We explored a covariant and separable form factor to construct a covariant chiral nucleon-nucleon interaction and
we found that a slightly better description of the $^1 S_0$
phase shifts  can be achieved by using the covariant form factor.
The resulting scattering length and effective range are in good agreement with the empirical values of the Nijmegen PWA.
In addition, a simultaneous description of the zero amplitude and the very shallow virtual bound state is obtained
at leading order, which
indicates that the relativistic $^1S_0$ potential is more consistent with the basic principle of EFTs, namely, being
able to describe the several typical small scale quantities simultaneously
at leading order. It should be noted that  the leading order relativistic chiral  description of the $^1S_0$ phenomena, as well as
the $J\le1$ partial wave phase shifts,
is qualitatively similar whether the covariant or the traditional separable form factor is adopted. It will be interesting to check what happens  at higher chiral orders.

\begin{acknowledgments}
X.-L.R. acknowledges the fruitful discussions with Prof. Bingwei Long.
K.-W.L. acknowledges financial support from the China Scholarship Council.
  This work was supported in part by the
 National Natural Science Foundation of China under Grants  No. 11375024, No. 11522539, No.  11735003,  No. 11335002,
 and No. 11775099, by NSFC and DFG  through funds provided to the
Sino-German CRC 110 ``Symmetries and the Emergence of Structure in QCD'' (NSFC Grant
No. 11621131001, DFG Grant No. TRR110),  the China
Postdoctoral Science Foundation under Grants No. 2016M600845, No. 2017T100008,
 the Major State 973 Program of China under Grant No. 2013CB834400, the Fundamental Research Funds for the
Central Universities.
\end{acknowledgments}

\bibliography{refs}

\begin{thebibliography}{90}
\expandafter\ifx\csname natexlab\endcsname\relax\def\natexlab#1{#1}\fi
\expandafter\ifx\csname bibnamefont\endcsname\relax
  \def\bibnamefont#1{#1}\fi
\expandafter\ifx\csname bibfnamefont\endcsname\relax
  \def\bibfnamefont#1{#1}\fi
\expandafter\ifx\csname citenamefont\endcsname\relax
  \def\citenamefont#1{#1}\fi
\expandafter\ifx\csname url\endcsname\relax
  \def\url#1{\texttt{#1}}\fi
\expandafter\ifx\csname urlprefix\endcsname\relax\def\urlprefix{URL }\fi
\providecommand{\bibinfo}[2]{#2}
\providecommand{\eprint}[2][]{\url{#2}}

\bibitem[{\citenamefont{Bedaque and van Kolck}(2002)}]{Bedaque:2002mn}
\bibinfo{author}{\bibfnamefont{P.~F.} \bibnamefont{Bedaque}} \bibnamefont{and}
  \bibinfo{author}{\bibfnamefont{U.}~\bibnamefont{van Kolck}},
  \bibinfo{journal}{Ann. Rev. Nucl. Part. Sci.} \textbf{\bibinfo{volume}{52}},
  \bibinfo{pages}{339} (\bibinfo{year}{2002}).

\bibitem[{\citenamefont{Epelbaum et~al.}(2009)\citenamefont{Epelbaum, Hammer,
  and Mei{\ss}ner}}]{Epelbaum:2008ga}
\bibinfo{author}{\bibfnamefont{E.}~\bibnamefont{Epelbaum}},
  \bibinfo{author}{\bibfnamefont{H.-W.} \bibnamefont{Hammer}},
  \bibnamefont{and} \bibinfo{author}{\bibfnamefont{U.-G.}
  \bibnamefont{Mei{\ss}ner}}, \bibinfo{journal}{Rev. Mod. Phys.}
  \textbf{\bibinfo{volume}{81}}, \bibinfo{pages}{1773} (\bibinfo{year}{2009}).

\bibitem[{\citenamefont{Machleidt and Entem}(2011)}]{Machleidt:2011zz}
\bibinfo{author}{\bibfnamefont{R.}~\bibnamefont{Machleidt}} \bibnamefont{and}
  \bibinfo{author}{\bibfnamefont{D.~R.} \bibnamefont{Entem}},
  \bibinfo{journal}{Phys. Rept.} \textbf{\bibinfo{volume}{503}},
  \bibinfo{pages}{1} (\bibinfo{year}{2011}).

\bibitem[{\citenamefont{Epelbaum and Mei{\ss}ner}(2012)}]{Epelbaum:2012vx}
\bibinfo{author}{\bibfnamefont{E.}~\bibnamefont{Epelbaum}} \bibnamefont{and}
  \bibinfo{author}{\bibfnamefont{U.-G.} \bibnamefont{Mei{\ss}ner}},
  \bibinfo{journal}{Ann. Rev. Nucl. Part. Sci.} \textbf{\bibinfo{volume}{62}},
  \bibinfo{pages}{159} (\bibinfo{year}{2012}).

\bibitem[{\citenamefont{Weinberg}(1979)}]{Weinberg:1978kz}
\bibinfo{author}{\bibfnamefont{S.}~\bibnamefont{Weinberg}},
  \bibinfo{journal}{Physica} \textbf{\bibinfo{volume}{A96}},
  \bibinfo{pages}{327} (\bibinfo{year}{1979}).

\bibitem[{\citenamefont{Weinberg}(1990)}]{Weinberg:1990rz}
\bibinfo{author}{\bibfnamefont{S.}~\bibnamefont{Weinberg}},
  \bibinfo{journal}{Phys. Lett.} \textbf{\bibinfo{volume}{B251}},
  \bibinfo{pages}{288} (\bibinfo{year}{1990}).

\bibitem[{\citenamefont{Weinberg}(1991)}]{Weinberg:1991um}
\bibinfo{author}{\bibfnamefont{S.}~\bibnamefont{Weinberg}},
  \bibinfo{journal}{Nucl. Phys.} \textbf{\bibinfo{volume}{B363}},
  \bibinfo{pages}{3} (\bibinfo{year}{1991}).

\bibitem[{\citenamefont{Entem and Machleidt}(2003)}]{Entem:2003ft}
\bibinfo{author}{\bibfnamefont{D.~R.} \bibnamefont{Entem}} \bibnamefont{and}
  \bibinfo{author}{\bibfnamefont{R.}~\bibnamefont{Machleidt}},
  \bibinfo{journal}{Phys. Rev.} \textbf{\bibinfo{volume}{C68}},
  \bibinfo{pages}{041001} (\bibinfo{year}{2003}).

\bibitem[{\citenamefont{Epelbaum et~al.}(2005)\citenamefont{Epelbaum, Glockle,
  and Mei{\ss}ner}}]{Epelbaum:2004fk}
\bibinfo{author}{\bibfnamefont{E.}~\bibnamefont{Epelbaum}},
  \bibinfo{author}{\bibfnamefont{W.}~\bibnamefont{Glockle}}, \bibnamefont{and}
  \bibinfo{author}{\bibfnamefont{U.-G.} \bibnamefont{Mei{\ss}ner}},
  \bibinfo{journal}{Nucl. Phys.} \textbf{\bibinfo{volume}{A747}},
  \bibinfo{pages}{362} (\bibinfo{year}{2005}).

\bibitem[{\citenamefont{Epelbaum et~al.}(2011)\citenamefont{Epelbaum, Krebs,
  Lee, and Mei{\ss}ner}}]{Epelbaum:2011md}
\bibinfo{author}{\bibfnamefont{E.}~\bibnamefont{Epelbaum}},
  \bibinfo{author}{\bibfnamefont{H.}~\bibnamefont{Krebs}},
  \bibinfo{author}{\bibfnamefont{D.}~\bibnamefont{Lee}}, \bibnamefont{and}
  \bibinfo{author}{\bibfnamefont{U.-G.} \bibnamefont{Mei{\ss}ner}},
  \bibinfo{journal}{Phys. Rev. Lett.} \textbf{\bibinfo{volume}{106}},
  \bibinfo{pages}{192501} (\bibinfo{year}{2011}).

\bibitem[{\citenamefont{Tews et~al.}(2013)\citenamefont{Tews, Kr\"{u}ger,
  Hebeler, and Schwenk}}]{Tews:2012fj}
\bibinfo{author}{\bibfnamefont{I.}~\bibnamefont{Tews}},
  \bibinfo{author}{\bibfnamefont{T.}~\bibnamefont{Kr\"{u}ger}},
  \bibinfo{author}{\bibfnamefont{K.}~\bibnamefont{Hebeler}}, \bibnamefont{and}
  \bibinfo{author}{\bibfnamefont{A.}~\bibnamefont{Schwenk}},
  \bibinfo{journal}{Phys. Rev. Lett.} \textbf{\bibinfo{volume}{110}},
  \bibinfo{pages}{032504} (\bibinfo{year}{2013}).

\bibitem[{\citenamefont{Epelbaum et~al.}(2013)\citenamefont{Epelbaum, Krebs,
  L\"{a}hde, Lee, and Mei{\ss}ner}}]{Epelbaum:2012iu}
\bibinfo{author}{\bibfnamefont{E.}~\bibnamefont{Epelbaum}},
  \bibinfo{author}{\bibfnamefont{H.}~\bibnamefont{Krebs}},
  \bibinfo{author}{\bibfnamefont{T.~A.} \bibnamefont{L\"{a}hde}},
  \bibinfo{author}{\bibfnamefont{D.}~\bibnamefont{Lee}}, \bibnamefont{and}
  \bibinfo{author}{\bibfnamefont{U.-G.} \bibnamefont{Mei{\ss}ner}},
  \bibinfo{journal}{Phys. Rev. Lett.} \textbf{\bibinfo{volume}{110}},
  \bibinfo{pages}{112502} (\bibinfo{year}{2013}).

\bibitem[{\citenamefont{Hergert
  et~al.}(2013{\natexlab{a}})\citenamefont{Hergert, Bogner, Binder, Calci,
  Langhammer, Roth, and Schwenk}}]{Hergert:2012nb}
\bibinfo{author}{\bibfnamefont{H.}~\bibnamefont{Hergert}},
  \bibinfo{author}{\bibfnamefont{S.~K.} \bibnamefont{Bogner}},
  \bibinfo{author}{\bibfnamefont{S.}~\bibnamefont{Binder}},
  \bibinfo{author}{\bibfnamefont{A.}~\bibnamefont{Calci}},
  \bibinfo{author}{\bibfnamefont{J.}~\bibnamefont{Langhammer}},
  \bibinfo{author}{\bibfnamefont{R.}~\bibnamefont{Roth}}, \bibnamefont{and}
  \bibinfo{author}{\bibfnamefont{A.}~\bibnamefont{Schwenk}},
  \bibinfo{journal}{Phys. Rev.} \textbf{\bibinfo{volume}{C87}},
  \bibinfo{pages}{034307} (\bibinfo{year}{2013}{\natexlab{a}}).

\bibitem[{\citenamefont{Hergert
  et~al.}(2013{\natexlab{b}})\citenamefont{Hergert, Binder, Calci, Langhammer,
  and Roth}}]{Hergert:2013vag}
\bibinfo{author}{\bibfnamefont{H.}~\bibnamefont{Hergert}},
  \bibinfo{author}{\bibfnamefont{S.}~\bibnamefont{Binder}},
  \bibinfo{author}{\bibfnamefont{A.}~\bibnamefont{Calci}},
  \bibinfo{author}{\bibfnamefont{J.}~\bibnamefont{Langhammer}},
  \bibnamefont{and} \bibinfo{author}{\bibfnamefont{R.}~\bibnamefont{Roth}},
  \bibinfo{journal}{Phys. Rev. Lett.} \textbf{\bibinfo{volume}{110}},
  \bibinfo{pages}{242501} (\bibinfo{year}{2013}{\natexlab{b}}).

\bibitem[{\citenamefont{Epelbaum et~al.}(2014)\citenamefont{Epelbaum, Krebs,
  L\"{a}hde, Lee, Mei{\ss}ner, and Rupak}}]{Epelbaum:2013paa}
\bibinfo{author}{\bibfnamefont{E.}~\bibnamefont{Epelbaum}},
  \bibinfo{author}{\bibfnamefont{H.}~\bibnamefont{Krebs}},
  \bibinfo{author}{\bibfnamefont{T.~A.} \bibnamefont{L\"{a}hde}},
  \bibinfo{author}{\bibfnamefont{D.}~\bibnamefont{Lee}},
  \bibinfo{author}{\bibfnamefont{U.-G.} \bibnamefont{Mei{\ss}ner}},
  \bibnamefont{and} \bibinfo{author}{\bibfnamefont{G.}~\bibnamefont{Rupak}},
  \bibinfo{journal}{Phys. Rev. Lett.} \textbf{\bibinfo{volume}{112}},
  \bibinfo{pages}{102501} (\bibinfo{year}{2014}).

\bibitem[{\citenamefont{Jansen et~al.}(2014)\citenamefont{Jansen, Engel, Hagen,
  Navratil, and Signoracci}}]{Jansen:2014qxa}
\bibinfo{author}{\bibfnamefont{G.~R.} \bibnamefont{Jansen}},
  \bibinfo{author}{\bibfnamefont{J.}~\bibnamefont{Engel}},
  \bibinfo{author}{\bibfnamefont{G.}~\bibnamefont{Hagen}},
  \bibinfo{author}{\bibfnamefont{P.}~\bibnamefont{Navratil}}, \bibnamefont{and}
  \bibinfo{author}{\bibfnamefont{A.}~\bibnamefont{Signoracci}},
  \bibinfo{journal}{Phys. Rev. Lett.} \textbf{\bibinfo{volume}{113}},
  \bibinfo{pages}{142502} (\bibinfo{year}{2014}).

\bibitem[{\citenamefont{Bogner et~al.}(2014)\citenamefont{Bogner, Hergert,
  Holt, Schwenk, Binder, Calci, Langhammer, and Roth}}]{Bogner:2014baa}
\bibinfo{author}{\bibfnamefont{S.~K.} \bibnamefont{Bogner}},
  \bibinfo{author}{\bibfnamefont{H.}~\bibnamefont{Hergert}},
  \bibinfo{author}{\bibfnamefont{J.~D.} \bibnamefont{Holt}},
  \bibinfo{author}{\bibfnamefont{A.}~\bibnamefont{Schwenk}},
  \bibinfo{author}{\bibfnamefont{S.}~\bibnamefont{Binder}},
  \bibinfo{author}{\bibfnamefont{A.}~\bibnamefont{Calci}},
  \bibinfo{author}{\bibfnamefont{J.}~\bibnamefont{Langhammer}},
  \bibnamefont{and} \bibinfo{author}{\bibfnamefont{R.}~\bibnamefont{Roth}},
  \bibinfo{journal}{Phys. Rev. Lett.} \textbf{\bibinfo{volume}{113}},
  \bibinfo{pages}{142501} (\bibinfo{year}{2014}).

\bibitem[{\citenamefont{Lynn et~al.}(2014)\citenamefont{Lynn, Carlson,
  Epelbaum, Gandolfi, Gezerlis, and Schwenk}}]{Lynn:2014zia}
\bibinfo{author}{\bibfnamefont{J.~E.} \bibnamefont{Lynn}},
  \bibinfo{author}{\bibfnamefont{J.}~\bibnamefont{Carlson}},
  \bibinfo{author}{\bibfnamefont{E.}~\bibnamefont{Epelbaum}},
  \bibinfo{author}{\bibfnamefont{S.}~\bibnamefont{Gandolfi}},
  \bibinfo{author}{\bibfnamefont{A.}~\bibnamefont{Gezerlis}}, \bibnamefont{and}
  \bibinfo{author}{\bibfnamefont{A.}~\bibnamefont{Schwenk}},
  \bibinfo{journal}{Phys. Rev. Lett.} \textbf{\bibinfo{volume}{113}},
  \bibinfo{pages}{192501} (\bibinfo{year}{2014}).

\bibitem[{\citenamefont{Hagen et~al.}(2015)}]{Hagen:2015yea}
\bibinfo{author}{\bibfnamefont{G.}~\bibnamefont{Hagen}} \bibnamefont{et~al.},
  \bibinfo{journal}{Nature Phys.} \textbf{\bibinfo{volume}{12}},
  \bibinfo{pages}{186} (\bibinfo{year}{2015}).

\bibitem[{\citenamefont{Elhatisari et~al.}(2015)\citenamefont{Elhatisari, Lee,
  Rupak, Epelbaum, Krebs, L\"{a}hde, Luu, and
  Mei{\ss}ner}}]{Elhatisari:2015iga}
\bibinfo{author}{\bibfnamefont{S.}~\bibnamefont{Elhatisari}},
  \bibinfo{author}{\bibfnamefont{D.}~\bibnamefont{Lee}},
  \bibinfo{author}{\bibfnamefont{G.}~\bibnamefont{Rupak}},
  \bibinfo{author}{\bibfnamefont{E.}~\bibnamefont{Epelbaum}},
  \bibinfo{author}{\bibfnamefont{H.}~\bibnamefont{Krebs}},
  \bibinfo{author}{\bibfnamefont{T.~A.} \bibnamefont{L\"{a}hde}},
  \bibinfo{author}{\bibfnamefont{T.}~\bibnamefont{Luu}}, \bibnamefont{and}
  \bibinfo{author}{\bibfnamefont{U.-G.} \bibnamefont{Mei{\ss}ner}},
  \bibinfo{journal}{Nature} \textbf{\bibinfo{volume}{528}},
  \bibinfo{pages}{111} (\bibinfo{year}{2015}).

\bibitem[{\citenamefont{Lapoux et~al.}(2016)\citenamefont{Lapoux, Som\`{a},
  Barbieri, Hergert, Holt, and Stroberg}}]{Lapoux:2016exf}
\bibinfo{author}{\bibfnamefont{V.}~\bibnamefont{Lapoux}},
  \bibinfo{author}{\bibfnamefont{V.}~\bibnamefont{Som\`{a}}},
  \bibinfo{author}{\bibfnamefont{C.}~\bibnamefont{Barbieri}},
  \bibinfo{author}{\bibfnamefont{H.}~\bibnamefont{Hergert}},
  \bibinfo{author}{\bibfnamefont{J.~D.} \bibnamefont{Holt}}, \bibnamefont{and}
  \bibinfo{author}{\bibfnamefont{S.}~\bibnamefont{Stroberg}},
  \bibinfo{journal}{Phys. Rev. Lett.} \textbf{\bibinfo{volume}{117}},
  \bibinfo{pages}{052501} (\bibinfo{year}{2016}).

\bibitem[{\citenamefont{Elhatisari et~al.}(2017)\citenamefont{Elhatisari,
  Epelbaum, Krebs, L\"{a}hde, Lee, Li, Lu, Mei{\ss}ner, and
  Rupak}}]{Elhatisari:2017eno}
\bibinfo{author}{\bibfnamefont{S.}~\bibnamefont{Elhatisari}},
  \bibinfo{author}{\bibfnamefont{E.}~\bibnamefont{Epelbaum}},
  \bibinfo{author}{\bibfnamefont{H.}~\bibnamefont{Krebs}},
  \bibinfo{author}{\bibfnamefont{T.~A.} \bibnamefont{L\"{a}hde}},
  \bibinfo{author}{\bibfnamefont{D.}~\bibnamefont{Lee}},
  \bibinfo{author}{\bibfnamefont{N.}~\bibnamefont{Li}},
  \bibinfo{author}{\bibfnamefont{B.-n.} \bibnamefont{Lu}},
  \bibinfo{author}{\bibfnamefont{U.-G.} \bibnamefont{Mei{\ss}ner}},
  \bibnamefont{and} \bibinfo{author}{\bibfnamefont{G.}~\bibnamefont{Rupak}},
  \bibinfo{journal}{Phys. Rev. Lett.} \textbf{\bibinfo{volume}{119}},
  \bibinfo{pages}{222505} (\bibinfo{year}{2017}).

\bibitem[{\citenamefont{Epelbaum
  et~al.}(2015{\natexlab{a}})\citenamefont{Epelbaum, Krebs, and
  Mei{\ss}ner}}]{Epelbaum:2014sza}
\bibinfo{author}{\bibfnamefont{E.}~\bibnamefont{Epelbaum}},
  \bibinfo{author}{\bibfnamefont{H.}~\bibnamefont{Krebs}}, \bibnamefont{and}
  \bibinfo{author}{\bibfnamefont{U.-G.} \bibnamefont{Mei{\ss}ner}},
  \bibinfo{journal}{Phys. Rev. Lett.} \textbf{\bibinfo{volume}{115}},
  \bibinfo{pages}{122301} (\bibinfo{year}{2015}{\natexlab{a}}).

\bibitem[{\citenamefont{Reinert et~al.}(2018)\citenamefont{Reinert, Krebs, and
  Epelbaum}}]{Reinert:2017usi}
\bibinfo{author}{\bibfnamefont{P.}~\bibnamefont{Reinert}},
  \bibinfo{author}{\bibfnamefont{H.}~\bibnamefont{Krebs}}, \bibnamefont{and}
  \bibinfo{author}{\bibfnamefont{E.}~\bibnamefont{Epelbaum}},
  \bibinfo{journal}{Eur. Phys. J. A} \textbf{\bibinfo{volume}{54}},
  \bibinfo{pages}{86} (\bibinfo{year}{2018}), \eprint{1711.08821}.

\bibitem[{\citenamefont{Entem et~al.}(2015)\citenamefont{Entem, Kaiser,
  Machleidt, and Nosyk}}]{Entem:2014msa}
\bibinfo{author}{\bibfnamefont{D.~R.} \bibnamefont{Entem}},
  \bibinfo{author}{\bibfnamefont{N.}~\bibnamefont{Kaiser}},
  \bibinfo{author}{\bibfnamefont{R.}~\bibnamefont{Machleidt}},
  \bibnamefont{and} \bibinfo{author}{\bibfnamefont{Y.}~\bibnamefont{Nosyk}},
  \bibinfo{journal}{Phys. Rev.} \textbf{\bibinfo{volume}{C91}},
  \bibinfo{pages}{014002} (\bibinfo{year}{2015}).

\bibitem[{\citenamefont{Entem et~al.}(2017)\citenamefont{Entem, Machleidt, and
  Nosyk}}]{Entem:2017gor}
\bibinfo{author}{\bibfnamefont{D.~R.} \bibnamefont{Entem}},
  \bibinfo{author}{\bibfnamefont{R.}~\bibnamefont{Machleidt}},
  \bibnamefont{and} \bibinfo{author}{\bibfnamefont{Y.}~\bibnamefont{Nosyk}},
  \bibinfo{journal}{Phys. Rev.} \textbf{\bibinfo{volume}{C96}},
  \bibinfo{pages}{024004} (\bibinfo{year}{2017}).

\bibitem[{\citenamefont{Stoks et~al.}(1994)\citenamefont{Stoks, Klomp,
  Terheggen, and de~Swart}}]{Stoks:1994wp}
\bibinfo{author}{\bibfnamefont{V.~G.~J.} \bibnamefont{Stoks}},
  \bibinfo{author}{\bibfnamefont{R.~A.~M.} \bibnamefont{Klomp}},
  \bibinfo{author}{\bibfnamefont{C.~P.~F.} \bibnamefont{Terheggen}},
  \bibnamefont{and} \bibinfo{author}{\bibfnamefont{J.~J.}
  \bibnamefont{de~Swart}}, \bibinfo{journal}{Phys. Rev.}
  \textbf{\bibinfo{volume}{C49}}, \bibinfo{pages}{2950} (\bibinfo{year}{1994}).

\bibitem[{\citenamefont{Wiringa et~al.}(1995)\citenamefont{Wiringa, Stoks, and
  Schiavilla}}]{Wiringa:1994wb}
\bibinfo{author}{\bibfnamefont{R.~B.} \bibnamefont{Wiringa}},
  \bibinfo{author}{\bibfnamefont{V.~G.~J.} \bibnamefont{Stoks}},
  \bibnamefont{and}
  \bibinfo{author}{\bibfnamefont{R.}~\bibnamefont{Schiavilla}},
  \bibinfo{journal}{Phys. Rev.} \textbf{\bibinfo{volume}{C51}},
  \bibinfo{pages}{38} (\bibinfo{year}{1995}).

\bibitem[{\citenamefont{Machleidt}(2001)}]{Machleidt:2000ge}
\bibinfo{author}{\bibfnamefont{R.}~\bibnamefont{Machleidt}},
  \bibinfo{journal}{Phys. Rev.} \textbf{\bibinfo{volume}{C63}},
  \bibinfo{pages}{024001} (\bibinfo{year}{2001}).

\bibitem[{\citenamefont{Meng}(2016)}]{Mengbook}
\bibinfo{editor}{\bibfnamefont{J.}~\bibnamefont{Meng}}, ed.,
  \emph{\bibinfo{title}{In {\it Relativistic Density Functional for Nuclear
  Structure}, International Review of Nuclear Physics Vol. 10}}
  (\bibinfo{publisher}{World Scientific, Singapore}, \bibinfo{year}{2016}).

\bibitem[{\citenamefont{Geng et~al.}(2008)\citenamefont{Geng, Martin~Camalich,
  Alvarez-Ruso, and Vicente~Vacas}}]{Geng:2008mf}
\bibinfo{author}{\bibfnamefont{L.-S.} \bibnamefont{Geng}},
  \bibinfo{author}{\bibfnamefont{J.}~\bibnamefont{Martin~Camalich}},
  \bibinfo{author}{\bibfnamefont{L.}~\bibnamefont{Alvarez-Ruso}},
  \bibnamefont{and} \bibinfo{author}{\bibfnamefont{M.~J.}
  \bibnamefont{Vicente~Vacas}}, \bibinfo{journal}{Phys. Rev. Lett.}
  \textbf{\bibinfo{volume}{101}}, \bibinfo{pages}{222002}
  (\bibinfo{year}{2008}).

\bibitem[{\citenamefont{Ren et~al.}(2012)\citenamefont{Ren, Geng,
  Martin~Camalich, Meng, and Toki}}]{Ren:2012aj}
\bibinfo{author}{\bibfnamefont{X.-L.} \bibnamefont{Ren}},
  \bibinfo{author}{\bibfnamefont{L.-S.} \bibnamefont{Geng}},
  \bibinfo{author}{\bibfnamefont{J.}~\bibnamefont{Martin~Camalich}},
  \bibinfo{author}{\bibfnamefont{J.}~\bibnamefont{Meng}}, \bibnamefont{and}
  \bibinfo{author}{\bibfnamefont{H.}~\bibnamefont{Toki}},
  \bibinfo{journal}{JHEP} \textbf{\bibinfo{volume}{12}}, \bibinfo{pages}{073}
  (\bibinfo{year}{2012}).

\bibitem[{\citenamefont{Ren et~al.}(2015)\citenamefont{Ren, Geng, and
  Meng}}]{Ren:2014vea}
\bibinfo{author}{\bibfnamefont{X.-L.} \bibnamefont{Ren}},
  \bibinfo{author}{\bibfnamefont{L.-S.} \bibnamefont{Geng}}, \bibnamefont{and}
  \bibinfo{author}{\bibfnamefont{J.}~\bibnamefont{Meng}},
  \bibinfo{journal}{Phys. Rev.} \textbf{\bibinfo{volume}{D91}},
  \bibinfo{pages}{051502} (\bibinfo{year}{2015}).

\bibitem[{\citenamefont{Altenbuchinger
  et~al.}(2014)\citenamefont{Altenbuchinger, Geng, and
  Weise}}]{Altenbuchinger:2013vwa}
\bibinfo{author}{\bibfnamefont{M.}~\bibnamefont{Altenbuchinger}},
  \bibinfo{author}{\bibfnamefont{L.-S.} \bibnamefont{Geng}}, \bibnamefont{and}
  \bibinfo{author}{\bibfnamefont{W.}~\bibnamefont{Weise}},
  \bibinfo{journal}{Phys. Rev.} \textbf{\bibinfo{volume}{D89}},
  \bibinfo{pages}{014026} (\bibinfo{year}{2014}).

\bibitem[{\citenamefont{Geng}(2013)}]{Geng:2013xn}
\bibinfo{author}{\bibfnamefont{L.-S.} \bibnamefont{Geng}},
  \bibinfo{journal}{Front. Phys.(Beijing)} \textbf{\bibinfo{volume}{8}},
  \bibinfo{pages}{328} (\bibinfo{year}{2013}).

\bibitem[{\citenamefont{Ren et~al.}(2018)\citenamefont{Ren, Li, Geng, Long,
  Ring, and Meng}}]{Ren:2016jna}
\bibinfo{author}{\bibfnamefont{X.-L.} \bibnamefont{Ren}},
  \bibinfo{author}{\bibfnamefont{K.-W.} \bibnamefont{Li}},
  \bibinfo{author}{\bibfnamefont{L.-S.} \bibnamefont{Geng}},
  \bibinfo{author}{\bibfnamefont{B.-W.} \bibnamefont{Long}},
  \bibinfo{author}{\bibfnamefont{P.}~\bibnamefont{Ring}}, \bibnamefont{and}
  \bibinfo{author}{\bibfnamefont{J.}~\bibnamefont{Meng}},
  \bibinfo{journal}{Chin. Phys. C} \textbf{\bibinfo{volume}{42}},
  \bibinfo{pages}{014103} (\bibinfo{year}{2018}).

\bibitem[{\citenamefont{Li et~al.}(2018{\natexlab{a}})\citenamefont{Li, Ren,
  Geng, and Long}}]{Li:2016mln}
\bibinfo{author}{\bibfnamefont{K.-W.} \bibnamefont{Li}},
  \bibinfo{author}{\bibfnamefont{X.-L.} \bibnamefont{Ren}},
  \bibinfo{author}{\bibfnamefont{L.-S.} \bibnamefont{Geng}}, \bibnamefont{and}
  \bibinfo{author}{\bibfnamefont{B.-W.} \bibnamefont{Long}},
  \bibinfo{journal}{Chin. Phys. C} \textbf{\bibinfo{volume}{42}},
  \bibinfo{pages}{014105} (\bibinfo{year}{2018}{\natexlab{a}}),
  \eprint{1612.08482}.

\bibitem[{\citenamefont{Song et~al.}(2018)\citenamefont{Song, Li, and
  Geng}}]{Song:2018qqm}
\bibinfo{author}{\bibfnamefont{J.}~\bibnamefont{Song}},
  \bibinfo{author}{\bibfnamefont{K.-W.} \bibnamefont{Li}}, \bibnamefont{and}
  \bibinfo{author}{\bibfnamefont{L.-S.} \bibnamefont{Geng}},
  \bibinfo{journal}{Phys. Rev. C} \textbf{\bibinfo{volume}{97}},
  \bibinfo{pages}{065201} (\bibinfo{year}{2018}), \eprint{1802.04433}.

\bibitem[{\citenamefont{Li et~al.}(2018{\natexlab{b}})\citenamefont{Li, Hyodo,
  and Geng}}]{Li:2018tbt}
\bibinfo{author}{\bibfnamefont{K.-W.} \bibnamefont{Li}},
  \bibinfo{author}{\bibfnamefont{T.}~\bibnamefont{Hyodo}}, \bibnamefont{and}
  \bibinfo{author}{\bibfnamefont{L.-S.} \bibnamefont{Geng}},
  \bibinfo{journal}{Phys. Rev. C} \textbf{\bibinfo{volume}{98}},
  \bibinfo{pages}{065203} (\bibinfo{year}{2018}{\natexlab{b}}),
  \eprint{1809.03199}.

\bibitem[{\citenamefont{Xiao et~al.}(2019)\citenamefont{Xiao, Geng, and
  Ren}}]{Xiao:2018jot}
\bibinfo{author}{\bibfnamefont{Y.}~\bibnamefont{Xiao}},
  \bibinfo{author}{\bibfnamefont{L.-S.} \bibnamefont{Geng}}, \bibnamefont{and}
  \bibinfo{author}{\bibfnamefont{X.-L.} \bibnamefont{Ren}},
  \bibinfo{journal}{Phys. Rev. C} \textbf{\bibinfo{volume}{99}},
  \bibinfo{pages}{024004} (\bibinfo{year}{2019}), \eprint{1812.03005}.

\bibitem[{\citenamefont{Bai et~al.}(2020)\citenamefont{Bai, Wang, Xiao, and
  Geng}}]{Bai:2020yml}
\bibinfo{author}{\bibfnamefont{Q.-Q.} \bibnamefont{Bai}},
  \bibinfo{author}{\bibfnamefont{C.-X.} \bibnamefont{Wang}},
  \bibinfo{author}{\bibfnamefont{Y.}~\bibnamefont{Xiao}}, \bibnamefont{and}
  \bibinfo{author}{\bibfnamefont{L.-S.} \bibnamefont{Geng}},
  \bibinfo{journal}{Phys. Lett.} \textbf{\bibinfo{volume}{B}},
  \bibinfo{pages}{135745} (\bibinfo{year}{2020}), \eprint{2007.01638}.

\bibitem[{\citenamefont{Xiao et~al.}(2020)\citenamefont{Xiao, Wang, Lu, and
  Geng}}]{Xiao:2020ozd}
\bibinfo{author}{\bibfnamefont{Y.}~\bibnamefont{Xiao}},
  \bibinfo{author}{\bibfnamefont{C.-X.} \bibnamefont{Wang}},
  \bibinfo{author}{\bibfnamefont{J.-X.} \bibnamefont{Lu}}, \bibnamefont{and}
  \bibinfo{author}{\bibfnamefont{L.-S.} \bibnamefont{Geng}},
  \bibinfo{journal}{Phys. Rev. C} \textbf{\bibinfo{volume}{102}},
  \bibinfo{pages}{054001} (\bibinfo{year}{2020}), \eprint{2007.13675}.

\bibitem[{\citenamefont{Song et~al.}(2020)\citenamefont{Song, Xiao, Liu, Wang,
  Li, and Geng}}]{Song:2020isu}
\bibinfo{author}{\bibfnamefont{J.}~\bibnamefont{Song}},
  \bibinfo{author}{\bibfnamefont{Y.}~\bibnamefont{Xiao}},
  \bibinfo{author}{\bibfnamefont{Z.-W.} \bibnamefont{Liu}},
  \bibinfo{author}{\bibfnamefont{C.-X.} \bibnamefont{Wang}},
  \bibinfo{author}{\bibfnamefont{K.-W.} \bibnamefont{Li}}, \bibnamefont{and}
  \bibinfo{author}{\bibfnamefont{L.-S.} \bibnamefont{Geng}},
  \bibinfo{journal}{Phys. Rev. C} \textbf{\bibinfo{volume}{102}},
  \bibinfo{pages}{065208} (\bibinfo{year}{2020}), \eprint{2010.06916}.

\bibitem[{\citenamefont{Liu et~al.}(2021)\citenamefont{Liu, Song, Li, and
  Geng}}]{Liu:2020uxi}
\bibinfo{author}{\bibfnamefont{Z.-W.} \bibnamefont{Liu}},
  \bibinfo{author}{\bibfnamefont{J.}~\bibnamefont{Song}},
  \bibinfo{author}{\bibfnamefont{K.-W.} \bibnamefont{Li}}, \bibnamefont{and}
  \bibinfo{author}{\bibfnamefont{L.-S.} \bibnamefont{Geng}},
  \bibinfo{journal}{Phys. Rev. C} \textbf{\bibinfo{volume}{103}},
  \bibinfo{pages}{025201} (\bibinfo{year}{2021}), \eprint{2011.05510}.

\bibitem[{\citenamefont{Brockmann and Machleidt}(1990)}]{Brockmann:1990cn}
\bibinfo{author}{\bibfnamefont{R.}~\bibnamefont{Brockmann}} \bibnamefont{and}
  \bibinfo{author}{\bibfnamefont{R.}~\bibnamefont{Machleidt}},
  \bibinfo{journal}{Phys. Rev.} \textbf{\bibinfo{volume}{C42}},
  \bibinfo{pages}{1965} (\bibinfo{year}{1990}).

\bibitem[{\citenamefont{Shen et~al.}(2016)\citenamefont{Shen, Hu, Liang, Meng,
  Ring, and Zhang}}]{Shen:2016bva}
\bibinfo{author}{\bibfnamefont{S.}~\bibnamefont{Shen}},
  \bibinfo{author}{\bibfnamefont{J.}~\bibnamefont{Hu}},
  \bibinfo{author}{\bibfnamefont{H.}~\bibnamefont{Liang}},
  \bibinfo{author}{\bibfnamefont{J.}~\bibnamefont{Meng}},
  \bibinfo{author}{\bibfnamefont{P.}~\bibnamefont{Ring}}, \bibnamefont{and}
  \bibinfo{author}{\bibfnamefont{S.}~\bibnamefont{Zhang}},
  \bibinfo{journal}{Chin. Phys. Lett.} \textbf{\bibinfo{volume}{33}},
  \bibinfo{pages}{102103} (\bibinfo{year}{2016}).

\bibitem[{\citenamefont{Shen et~al.}(2017)\citenamefont{Shen, Liang, Meng,
  Ring, and Zhang}}]{Shen:2017vqr}
\bibinfo{author}{\bibfnamefont{S.}~\bibnamefont{Shen}},
  \bibinfo{author}{\bibfnamefont{H.}~\bibnamefont{Liang}},
  \bibinfo{author}{\bibfnamefont{J.}~\bibnamefont{Meng}},
  \bibinfo{author}{\bibfnamefont{P.}~\bibnamefont{Ring}}, \bibnamefont{and}
  \bibinfo{author}{\bibfnamefont{S.}~\bibnamefont{Zhang}},
  \bibinfo{journal}{Phys. Rev. C} \textbf{\bibinfo{volume}{96}},
  \bibinfo{pages}{014316} (\bibinfo{year}{2017}), \eprint{1705.01691}.

\bibitem[{\citenamefont{Shen et~al.}(2019)\citenamefont{Shen, Liang, Long,
  Meng, and Ring}}]{Shen:2019dls}
\bibinfo{author}{\bibfnamefont{S.}~\bibnamefont{Shen}},
  \bibinfo{author}{\bibfnamefont{H.}~\bibnamefont{Liang}},
  \bibinfo{author}{\bibfnamefont{W.~H.} \bibnamefont{Long}},
  \bibinfo{author}{\bibfnamefont{J.}~\bibnamefont{Meng}}, \bibnamefont{and}
  \bibinfo{author}{\bibfnamefont{P.}~\bibnamefont{Ring}},
  \bibinfo{journal}{Prog. Part. Nucl. Phys.} \textbf{\bibinfo{volume}{109}},
  \bibinfo{pages}{103713} (\bibinfo{year}{2019}), \eprint{1904.04977}.

\bibitem[{\citenamefont{Salpeter and Bethe}(1951)}]{Salpeter:1951sz}
\bibinfo{author}{\bibfnamefont{E.~E.} \bibnamefont{Salpeter}} \bibnamefont{and}
  \bibinfo{author}{\bibfnamefont{H.~A.} \bibnamefont{Bethe}},
  \bibinfo{journal}{Phys. Rev.} \textbf{\bibinfo{volume}{84}},
  \bibinfo{pages}{1232} (\bibinfo{year}{1951}).

\bibitem[{\citenamefont{Kadyshevsky}(1968)}]{Kadyshevsky:1967rs}
\bibinfo{author}{\bibfnamefont{V.~G.} \bibnamefont{Kadyshevsky}},
  \bibinfo{journal}{Nucl. Phys.} \textbf{\bibinfo{volume}{B6}},
  \bibinfo{pages}{125} (\bibinfo{year}{1968}).

\bibitem[{\citenamefont{Erkelenz}(1974)}]{erkelenz:1974uj}
\bibinfo{author}{\bibfnamefont{K.}~\bibnamefont{Erkelenz}},
  \bibinfo{journal}{Phys. Rept.} \textbf{\bibinfo{volume}{13}},
  \bibinfo{pages}{191} (\bibinfo{year}{1974}).

\bibitem[{\citenamefont{Ueda and Green}(1968)}]{Ueda:1969er}
\bibinfo{author}{\bibfnamefont{T.}~\bibnamefont{Ueda}} \bibnamefont{and}
  \bibinfo{author}{\bibfnamefont{A.~E.~S.} \bibnamefont{Green}},
  \bibinfo{journal}{Phys. Rev.} \textbf{\bibinfo{volume}{174}},
  \bibinfo{pages}{1304} (\bibinfo{year}{1968}).

\bibitem[{\citenamefont{Machleidt et~al.}(1987)\citenamefont{Machleidt,
  Holinde, and Elster}}]{Machleidt:1987hj}
\bibinfo{author}{\bibfnamefont{R.}~\bibnamefont{Machleidt}},
  \bibinfo{author}{\bibfnamefont{K.}~\bibnamefont{Holinde}}, \bibnamefont{and}
  \bibinfo{author}{\bibfnamefont{C.}~\bibnamefont{Elster}},
  \bibinfo{journal}{Phys. Rept.} \textbf{\bibinfo{volume}{149}},
  \bibinfo{pages}{1} (\bibinfo{year}{1987}).

\bibitem[{\citenamefont{Jackson et~al.}(1975)\citenamefont{Jackson, Riska, and
  Verwest}}]{Jackson:1975be}
\bibinfo{author}{\bibfnamefont{A.~D.} \bibnamefont{Jackson}},
  \bibinfo{author}{\bibfnamefont{D.~O.} \bibnamefont{Riska}}, \bibnamefont{and}
  \bibinfo{author}{\bibfnamefont{B.}~\bibnamefont{Verwest}},
  \bibinfo{journal}{Nucl. Phys.} \textbf{\bibinfo{volume}{A249}},
  \bibinfo{pages}{397} (\bibinfo{year}{1975}).

\bibitem[{\citenamefont{Holinde and Machleidt}(1976)}]{Holinde:1976qwa}
\bibinfo{author}{\bibfnamefont{K.}~\bibnamefont{Holinde}} \bibnamefont{and}
  \bibinfo{author}{\bibfnamefont{R.}~\bibnamefont{Machleidt}},
  \bibinfo{journal}{Nucl. Phys.} \textbf{\bibinfo{volume}{A256}},
  \bibinfo{pages}{479} (\bibinfo{year}{1976}).

\bibitem[{\citenamefont{Woloshyn and Jackson}(1972)}]{Woloshyn:1971zz}
\bibinfo{author}{\bibfnamefont{R.~M.} \bibnamefont{Woloshyn}} \bibnamefont{and}
  \bibinfo{author}{\bibfnamefont{A.~D.} \bibnamefont{Jackson}},
  \bibinfo{journal}{Nucl. Phys.} \textbf{\bibinfo{volume}{A185}},
  \bibinfo{pages}{131} (\bibinfo{year}{1972}).

\bibitem[{\citenamefont{Nagels et~al.}(1978)\citenamefont{Nagels, Rijken, and
  de~Swart}}]{Nagels:1977ze}
\bibinfo{author}{\bibfnamefont{M.~M.} \bibnamefont{Nagels}},
  \bibinfo{author}{\bibfnamefont{T.~A.} \bibnamefont{Rijken}},
  \bibnamefont{and} \bibinfo{author}{\bibfnamefont{J.~J.}
  \bibnamefont{de~Swart}}, \bibinfo{journal}{Phys. Rev.}
  \textbf{\bibinfo{volume}{D17}}, \bibinfo{pages}{768} (\bibinfo{year}{1978}).

\bibitem[{\citenamefont{Ordonez et~al.}(1994)\citenamefont{Ordonez, Ray, and
  van Kolck}}]{Ordonez:1993tn}
\bibinfo{author}{\bibfnamefont{C.}~\bibnamefont{Ordonez}},
  \bibinfo{author}{\bibfnamefont{L.}~\bibnamefont{Ray}}, \bibnamefont{and}
  \bibinfo{author}{\bibfnamefont{U.}~\bibnamefont{van Kolck}},
  \bibinfo{journal}{Phys. Rev. Lett.} \textbf{\bibinfo{volume}{72}},
  \bibinfo{pages}{1982} (\bibinfo{year}{1994}).

\bibitem[{\citenamefont{Epelbaum et~al.}(2000)\citenamefont{Epelbaum, Gloeckle,
  and Mei{\ss}ner}}]{Epelbaum:1999dj}
\bibinfo{author}{\bibfnamefont{E.}~\bibnamefont{Epelbaum}},
  \bibinfo{author}{\bibfnamefont{W.}~\bibnamefont{Gloeckle}}, \bibnamefont{and}
  \bibinfo{author}{\bibfnamefont{U.-G.} \bibnamefont{Mei{\ss}ner}},
  \bibinfo{journal}{Nucl. Phys.} \textbf{\bibinfo{volume}{A671}},
  \bibinfo{pages}{295} (\bibinfo{year}{2000}).

\bibitem[{\citenamefont{Nogga et~al.}(2005)\citenamefont{Nogga, Timmermans, and
  van Kolck}}]{Nogga:2005hy}
\bibinfo{author}{\bibfnamefont{A.}~\bibnamefont{Nogga}},
  \bibinfo{author}{\bibfnamefont{R.~G.~E.} \bibnamefont{Timmermans}},
  \bibnamefont{and} \bibinfo{author}{\bibfnamefont{U.}~\bibnamefont{van
  Kolck}}, \bibinfo{journal}{Phys. Rev.} \textbf{\bibinfo{volume}{C72}},
  \bibinfo{pages}{054006} (\bibinfo{year}{2005}).

\bibitem[{\citenamefont{Epelbaum}(2000)}]{Epelbaum:2000kv}
\bibinfo{author}{\bibfnamefont{E.}~\bibnamefont{Epelbaum}}, Ph.D. thesis,
  \bibinfo{school}{Julich, Forschungszentrum} (\bibinfo{year}{2000}).

\bibitem[{\citenamefont{Epelbaum
  et~al.}(2015{\natexlab{b}})\citenamefont{Epelbaum, Krebs, and
  Mei{\ss}ner}}]{Epelbaum:2014efa}
\bibinfo{author}{\bibfnamefont{E.}~\bibnamefont{Epelbaum}},
  \bibinfo{author}{\bibfnamefont{H.}~\bibnamefont{Krebs}}, \bibnamefont{and}
  \bibinfo{author}{\bibfnamefont{U.~G.} \bibnamefont{Mei{\ss}ner}},
  \bibinfo{journal}{Eur. Phys. J.} \textbf{\bibinfo{volume}{A51}},
  \bibinfo{pages}{53} (\bibinfo{year}{2015}{\natexlab{b}}).

\bibitem[{\citenamefont{van Kolck}(1999)}]{vanKolck:1998bw}
\bibinfo{author}{\bibfnamefont{U.}~\bibnamefont{van Kolck}},
  \bibinfo{journal}{Nucl. Phys.} \textbf{\bibinfo{volume}{A645}},
  \bibinfo{pages}{273} (\bibinfo{year}{1999}).

\bibitem[{\citenamefont{Lutz}(2000)}]{Lutz:1999yr}
\bibinfo{author}{\bibfnamefont{M.}~\bibnamefont{Lutz}}, \bibinfo{journal}{Nucl.
  Phys.} \textbf{\bibinfo{volume}{A677}}, \bibinfo{pages}{241}
  (\bibinfo{year}{2000}).

\bibitem[{\citenamefont{S\'anchez~S\'anchez
  et~al.}(2018)\citenamefont{S\'anchez~S\'anchez, Yang, Long, and van
  Kolck}}]{SanchezSanchez:2017tws}
\bibinfo{author}{\bibfnamefont{M.}~\bibnamefont{S\'anchez~S\'anchez}},
  \bibinfo{author}{\bibfnamefont{C.~J.} \bibnamefont{Yang}},
  \bibinfo{author}{\bibfnamefont{B.}~\bibnamefont{Long}}, \bibnamefont{and}
  \bibinfo{author}{\bibfnamefont{U.}~\bibnamefont{van Kolck}},
  \bibinfo{journal}{Phys. Rev. C} \textbf{\bibinfo{volume}{97}},
  \bibinfo{pages}{024001} (\bibinfo{year}{2018}), \eprint{1704.08524}.

\bibitem[{\citenamefont{Olive}(2016)}]{Olive:2016xmw}
\bibinfo{author}{\bibfnamefont{K.~A.} \bibnamefont{Olive}},
  \bibinfo{journal}{Chin. Phys.} \textbf{\bibinfo{volume}{C40}},
  \bibinfo{pages}{100001} (\bibinfo{year}{2016}).

\bibitem[{\citenamefont{Polinder et~al.}(2006)\citenamefont{Polinder,
  Haidenbauer, and Mei{\ss}ner}}]{Polinder:2006zh}
\bibinfo{author}{\bibfnamefont{H.}~\bibnamefont{Polinder}},
  \bibinfo{author}{\bibfnamefont{J.}~\bibnamefont{Haidenbauer}},
  \bibnamefont{and} \bibinfo{author}{\bibfnamefont{U.-G.}
  \bibnamefont{Mei{\ss}ner}}, \bibinfo{journal}{Nucl. Phys.}
  \textbf{\bibinfo{volume}{A779}}, \bibinfo{pages}{244} (\bibinfo{year}{2006}).

\bibitem[{\citenamefont{Djukanovic et~al.}(2007)\citenamefont{Djukanovic,
  Gegelia, Scherer, and Schindler}}]{Djukanovic:2007zz}
\bibinfo{author}{\bibfnamefont{D.}~\bibnamefont{Djukanovic}},
  \bibinfo{author}{\bibfnamefont{J.}~\bibnamefont{Gegelia}},
  \bibinfo{author}{\bibfnamefont{S.}~\bibnamefont{Scherer}}, \bibnamefont{and}
  \bibinfo{author}{\bibfnamefont{M.~R.} \bibnamefont{Schindler}},
  \bibinfo{journal}{Few Body Syst.} \textbf{\bibinfo{volume}{41}},
  \bibinfo{pages}{141} (\bibinfo{year}{2007}).

\bibitem[{\citenamefont{Woloshyn and Jackson}(1973)}]{Woloshyn:1974wm}
\bibinfo{author}{\bibfnamefont{R.~M.} \bibnamefont{Woloshyn}} \bibnamefont{and}
  \bibinfo{author}{\bibfnamefont{A.~D.} \bibnamefont{Jackson}},
  \bibinfo{journal}{Nucl. Phys.} \textbf{\bibinfo{volume}{B64}},
  \bibinfo{pages}{269} (\bibinfo{year}{1973}).

\bibitem[{\citenamefont{Thompson}(1970)}]{Thompson:1970wt}
\bibinfo{author}{\bibfnamefont{R.~H.} \bibnamefont{Thompson}},
  \bibinfo{journal}{Phys. Rev.} \textbf{\bibinfo{volume}{D1}},
  \bibinfo{pages}{110} (\bibinfo{year}{1970}).

\bibitem[{\citenamefont{Blankenbecler and Sugar}(1966)}]{Blankenbecler:1965gx}
\bibinfo{author}{\bibfnamefont{R.}~\bibnamefont{Blankenbecler}}
  \bibnamefont{and} \bibinfo{author}{\bibfnamefont{R.}~\bibnamefont{Sugar}},
  \bibinfo{journal}{Phys. Rev.} \textbf{\bibinfo{volume}{142}},
  \bibinfo{pages}{1051} (\bibinfo{year}{1966}).

\bibitem[{\citenamefont{Wang et~al.}(2021)\citenamefont{Wang, Geng, and
  Long}}]{Wang:2020myr}
\bibinfo{author}{\bibfnamefont{C.-X.} \bibnamefont{Wang}},
  \bibinfo{author}{\bibfnamefont{L.-S.} \bibnamefont{Geng}}, \bibnamefont{and}
  \bibinfo{author}{\bibfnamefont{B.}~\bibnamefont{Long}},
  \bibinfo{journal}{Chin. Phys. C} \textbf{\bibinfo{volume}{45}},
  \bibinfo{pages}{054101} (\bibinfo{year}{2021}), \eprint{2001.08483}.

\bibitem[{\citenamefont{Stoks et~al.}(1993)\citenamefont{Stoks, Klomp,
  Rentmeester, and de~Swart}}]{Stoks:1993tb}
\bibinfo{author}{\bibfnamefont{V.~G.~J.} \bibnamefont{Stoks}},
  \bibinfo{author}{\bibfnamefont{R.~A.~M.} \bibnamefont{Klomp}},
  \bibinfo{author}{\bibfnamefont{M.~C.~M.} \bibnamefont{Rentmeester}},
  \bibnamefont{and} \bibinfo{author}{\bibfnamefont{J.~J.}
  \bibnamefont{de~Swart}}, \bibinfo{journal}{Phys. Rev.}
  \textbf{\bibinfo{volume}{C48}}, \bibinfo{pages}{792} (\bibinfo{year}{1993}).

\bibitem[{\citenamefont{NN-OnLine}()}]{nnonline}
\bibinfo{author}{\bibnamefont{NN-OnLine}},
  \bibinfo{note}{http://nn-online.org}.

\bibitem[{\citenamefont{Pavon~Valderrama and
  Ruiz~Arriola}(2005)}]{PavonValderrama:2005ku}
\bibinfo{author}{\bibfnamefont{M.}~\bibnamefont{Pavon~Valderrama}}
  \bibnamefont{and}
  \bibinfo{author}{\bibfnamefont{E.}~\bibnamefont{Ruiz~Arriola}},
  \bibinfo{journal}{Phys. Rev.} \textbf{\bibinfo{volume}{C72}},
  \bibinfo{pages}{044007} (\bibinfo{year}{2005}).

\bibitem[{\citenamefont{Epelbaum and Gegelia}(2012)}]{Epelbaum:2012ua}
\bibinfo{author}{\bibfnamefont{E.}~\bibnamefont{Epelbaum}} \bibnamefont{and}
  \bibinfo{author}{\bibfnamefont{J.}~\bibnamefont{Gegelia}},
  \bibinfo{journal}{Phys. Lett.} \textbf{\bibinfo{volume}{B716}},
  \bibinfo{pages}{338} (\bibinfo{year}{2012}).

\bibitem[{\citenamefont{Kaplan et~al.}(1998)\citenamefont{Kaplan, Savage, and
  Wise}}]{Kaplan:1998tg}
\bibinfo{author}{\bibfnamefont{D.~B.} \bibnamefont{Kaplan}},
  \bibinfo{author}{\bibfnamefont{M.~J.} \bibnamefont{Savage}},
  \bibnamefont{and} \bibinfo{author}{\bibfnamefont{M.~B.} \bibnamefont{Wise}},
  \bibinfo{journal}{Phys. Lett.} \textbf{\bibinfo{volume}{B424}},
  \bibinfo{pages}{390} (\bibinfo{year}{1998}).

\bibitem[{\citenamefont{Birse}(2006)}]{Birse:2005um}
\bibinfo{author}{\bibfnamefont{M.~C.} \bibnamefont{Birse}},
  \bibinfo{journal}{Phys. Rev.} \textbf{\bibinfo{volume}{C74}},
  \bibinfo{pages}{014003} (\bibinfo{year}{2006}).

\bibitem[{\citenamefont{Timoteo et~al.}(2005)\citenamefont{Timoteo, Frederico,
  Delfino, and Tomio}}]{Timoteo:2005ia}
\bibinfo{author}{\bibfnamefont{V.~S.} \bibnamefont{Timoteo}},
  \bibinfo{author}{\bibfnamefont{T.}~\bibnamefont{Frederico}},
  \bibinfo{author}{\bibfnamefont{A.}~\bibnamefont{Delfino}}, \bibnamefont{and}
  \bibinfo{author}{\bibfnamefont{L.}~\bibnamefont{Tomio}},
  \bibinfo{journal}{Phys. Lett.} \textbf{\bibinfo{volume}{B621}},
  \bibinfo{pages}{109} (\bibinfo{year}{2005}).

\bibitem[{\citenamefont{Birse}(2007)}]{Birse:2007sx}
\bibinfo{author}{\bibfnamefont{M.~C.} \bibnamefont{Birse}},
  \bibinfo{journal}{Phys. Rev.} \textbf{\bibinfo{volume}{C76}},
  \bibinfo{pages}{034002} (\bibinfo{year}{2007}).

\bibitem[{\citenamefont{Yang et~al.}(2008)\citenamefont{Yang, Elster, and
  Phillips}}]{Yang:2007hb}
\bibinfo{author}{\bibfnamefont{C.~J.} \bibnamefont{Yang}},
  \bibinfo{author}{\bibfnamefont{C.}~\bibnamefont{Elster}}, \bibnamefont{and}
  \bibinfo{author}{\bibfnamefont{D.~R.} \bibnamefont{Phillips}},
  \bibinfo{journal}{Phys. Rev.} \textbf{\bibinfo{volume}{C77}},
  \bibinfo{pages}{014002} (\bibinfo{year}{2008}).

\bibitem[{\citenamefont{Yang et~al.}(2009{\natexlab{a}})\citenamefont{Yang,
  Elster, and Phillips}}]{Yang:2009kx}
\bibinfo{author}{\bibfnamefont{C.~J.} \bibnamefont{Yang}},
  \bibinfo{author}{\bibfnamefont{C.}~\bibnamefont{Elster}}, \bibnamefont{and}
  \bibinfo{author}{\bibfnamefont{D.~R.} \bibnamefont{Phillips}},
  \bibinfo{journal}{Phys. Rev.} \textbf{\bibinfo{volume}{C80}},
  \bibinfo{pages}{034002} (\bibinfo{year}{2009}{\natexlab{a}}).

\bibitem[{\citenamefont{Yang et~al.}(2009{\natexlab{b}})\citenamefont{Yang,
  Elster, and Phillips}}]{Yang:2009pn}
\bibinfo{author}{\bibfnamefont{C.~J.} \bibnamefont{Yang}},
  \bibinfo{author}{\bibfnamefont{C.}~\bibnamefont{Elster}}, \bibnamefont{and}
  \bibinfo{author}{\bibfnamefont{D.~R.} \bibnamefont{Phillips}},
  \bibinfo{journal}{Phys. Rev.} \textbf{\bibinfo{volume}{C80}},
  \bibinfo{pages}{044002} (\bibinfo{year}{2009}{\natexlab{b}}).

\bibitem[{\citenamefont{Valderrama}(2011)}]{Valderrama:2009ei}
\bibinfo{author}{\bibfnamefont{M.~P.} \bibnamefont{Valderrama}},
  \bibinfo{journal}{Phys. Rev.} \textbf{\bibinfo{volume}{C83}},
  \bibinfo{pages}{024003} (\bibinfo{year}{2011}).

\bibitem[{\citenamefont{Pavon~Valderrama}(2011)}]{Valderrama:2011mv}
\bibinfo{author}{\bibfnamefont{M.}~\bibnamefont{Pavon~Valderrama}},
  \bibinfo{journal}{Phys. Rev.} \textbf{\bibinfo{volume}{C84}},
  \bibinfo{pages}{064002} (\bibinfo{year}{2011}).

\bibitem[{\citenamefont{Long and Yang}(2011)}]{Long:2011qx}
\bibinfo{author}{\bibfnamefont{B.}~\bibnamefont{Long}} \bibnamefont{and}
  \bibinfo{author}{\bibfnamefont{C.~J.} \bibnamefont{Yang}},
  \bibinfo{journal}{Phys. Rev.} \textbf{\bibinfo{volume}{C84}},
  \bibinfo{pages}{057001} (\bibinfo{year}{2011}).

\bibitem[{\citenamefont{Long and Yang}(2012)}]{Long:2012ve}
\bibinfo{author}{\bibfnamefont{B.}~\bibnamefont{Long}} \bibnamefont{and}
  \bibinfo{author}{\bibfnamefont{C.~J.} \bibnamefont{Yang}},
  \bibinfo{journal}{Phys. Rev.} \textbf{\bibinfo{volume}{C86}},
  \bibinfo{pages}{024001} (\bibinfo{year}{2012}).

\bibitem[{\citenamefont{Epelbaum
  et~al.}(2015{\natexlab{c}})\citenamefont{Epelbaum, Gasparyan, Gegelia, and
  Krebs}}]{Epelbaum:2015sha}
\bibinfo{author}{\bibfnamefont{E.}~\bibnamefont{Epelbaum}},
  \bibinfo{author}{\bibfnamefont{A.~M.} \bibnamefont{Gasparyan}},
  \bibinfo{author}{\bibfnamefont{J.}~\bibnamefont{Gegelia}}, \bibnamefont{and}
  \bibinfo{author}{\bibfnamefont{H.}~\bibnamefont{Krebs}},
  \bibinfo{journal}{Eur. Phys. J.} \textbf{\bibinfo{volume}{A51}},
  \bibinfo{pages}{71} (\bibinfo{year}{2015}{\natexlab{c}}).

\bibitem[{\citenamefont{Baru et~al.}(2019)\citenamefont{Baru, Epelbaum,
  Gegelia, and Ren}}]{Baru:2019ndr}
\bibinfo{author}{\bibfnamefont{V.}~\bibnamefont{Baru}},
  \bibinfo{author}{\bibfnamefont{E.}~\bibnamefont{Epelbaum}},
  \bibinfo{author}{\bibfnamefont{J.}~\bibnamefont{Gegelia}}, \bibnamefont{and}
  \bibinfo{author}{\bibfnamefont{X.~L.} \bibnamefont{Ren}},
  \bibinfo{journal}{Phys. Lett.} \textbf{\bibinfo{volume}{B798}},
  \bibinfo{pages}{134987} (\bibinfo{year}{2019}), \eprint{1905.02116}.

\bibitem[{\citenamefont{Ren et~al.}(2020)\citenamefont{Ren, Epelbaum, and
  Gegelia}}]{Ren:2019qow}
\bibinfo{author}{\bibfnamefont{X.~L.} \bibnamefont{Ren}},
  \bibinfo{author}{\bibfnamefont{E.}~\bibnamefont{Epelbaum}}, \bibnamefont{and}
  \bibinfo{author}{\bibfnamefont{J.}~\bibnamefont{Gegelia}},
  \bibinfo{journal}{Phys. Rev.} \textbf{\bibinfo{volume}{C101}},
  \bibinfo{pages}{034001} (\bibinfo{year}{2020}), \eprint{1911.05616}.

\end{thebibliography}
%\begin{thebibliography}{99}
%
%\end{thebibliography}

\end{document}